\documentclass{elsarticle}

\usepackage{latexsym, amsmath, amsfonts, amssymb, graphicx, color, setspace}
\usepackage[latin1]{inputenc}

\date{}

\begin{document}

\parindent 0mm
\parskip   2mm
\renewcommand{\baselinestretch}{1.0}



\title{Opinion formation dynamics - Swift collective disillusionment triggered by unmet expectations}


\author[ktiethz]{Fariba Hashemi\corref{}}
\ead{fhashemi@ethz.ch}
\address[ktiethz]{ETH Zurich (ETHZ), Switzerland, and Karolinska Institutet, Sweden}
\author[unil]{Olivier Gallay\corref{cor}}
\ead{olivier.gallay@unil.ch}
\address[unil]{HEC, University of Lausanne, Switzerland}
\author[epfl]{Max-Olivier Hongler}
\ead{max.hongler@epfl.ch}
\address[epfl]{Ecole Polytechnique F\'ed\'erale de Lausanne (EPFL), Switzerland \\ \vspace{0.5cm} June 3 $^{\mbox{{\tiny rd}}}$, 2019}
\cortext[cor]{Corresponding author}



\begin{abstract}

We propose a microscopic model to describe how individual opinions shared between interacting agents initiate excessive collective expectations about a new idea or an innovation, followed by a swift collapse towards a dramatic collective disillusionment. The basic assumption which underlies the dynamics is that the information gathering process is not instantaneous but requires maturation. Agents steadily refine and update their personal opinion via a recurrent consultation of a public pool which stores information tokens (ITs). The expectation for the innovative idea is monitored in real-time by counting the number of stored ITs. The flow dynamics of ITs is assimilated to a single node queuing system (QS) with feedback loop. It incorporates the information pool (the waiting room), an IT inflow, and a service outflow that stylizes the information gathering process. Contrary to basic queuing theory, here the ITs roaming the QS are endowed with time-dependent internal variables. This additional dynamic information is used to construct the information maturation process. Such a maturation of the information introduces response delays into the dynamics, which ultimately generates the collective disillusionment trough. We illustrate the introduced generic modeling framework by considering in details the hype cycle dynamics, a key managerial topic when dealing with diffusion of innovation. In a second part of the paper, we introduce a stylized framework to detect, as soon as possible, the onset of the collective disillusionment phase, while minimizing the frequency of false alarms.

\end{abstract}



\maketitle
\thispagestyle{empty}
\vspace*{-2mm}


\section{Introduction}


\noindent  It is commonly observed that the spontaneous emergence of a buzz for innovative technologies, original fashion accessories, art works, and other novelties in general, is suddenly  cutoff  by an unpredictable, brisk collective disinterest or disillusionment. The  intrinsic uncertainty underlying such large amplitude oscillations generates an important source of risks for decision makers such as investors, producers, or consumers. In this paper, we model such enthusiasm and subsequent disinterest for an innovation as a macroscopic signature emerging from collective information exchanges within  a large swarm of autonomous agents, that are building individual opinions. The considered dynamics is based on the premise that agents jointly gather and enrich the content of a public pool of information. It yields a {\it stigmergic} mechanism, in which the collective opinion is self-generated via indirect interactions through repeated exchanges with the information pool. We also postulate that any information updating is basically ruled by an ageing process that enables opinions to gain maturity and relevance to progressively reveal potential shortcomings. This information ageing mechanism introduces delays into the dynamics, which is ultimately responsible for the emergence of large and swift collective declines. The present paper addresses the two following basic issues, namely: 
\begin{itemize}
  \item[(a)] How to stylize, from a multi-agent perspective, the emergence of such collective positive trends for an innovation, followed by sudden disinterest?
  \item[(b)] How to detect, as swiftly as possible and considering a given risk for error, the onset of the disinterest phase?
\end{itemize}
\noindent As a paradigmatic illustration of the proposed model, we elaborate further on the well-known technological innovation {\it qualitative hype cycle dynamics}, as sketched in Figure \ref{hype_cycle} below. 

\vspace{0.4cm}
\textit{The Hype Cycle Model}

The hype cycle model has been proposed by Gartner Inc. as a framework to facilitate the understanding and forecasting of the trajectories, maturity level, and market potential of early-stage technologies \cite{Fenn2008}. The framework provides a cross-industry perspective on technologies and trends, mapping the journey of new and emerging technologies through their evolution via distinct phases, as illustrated in Figure \ref{hype_cycle}. 
\begin{figure}[htbp]
\begin{center}
\includegraphics[height=8cm]{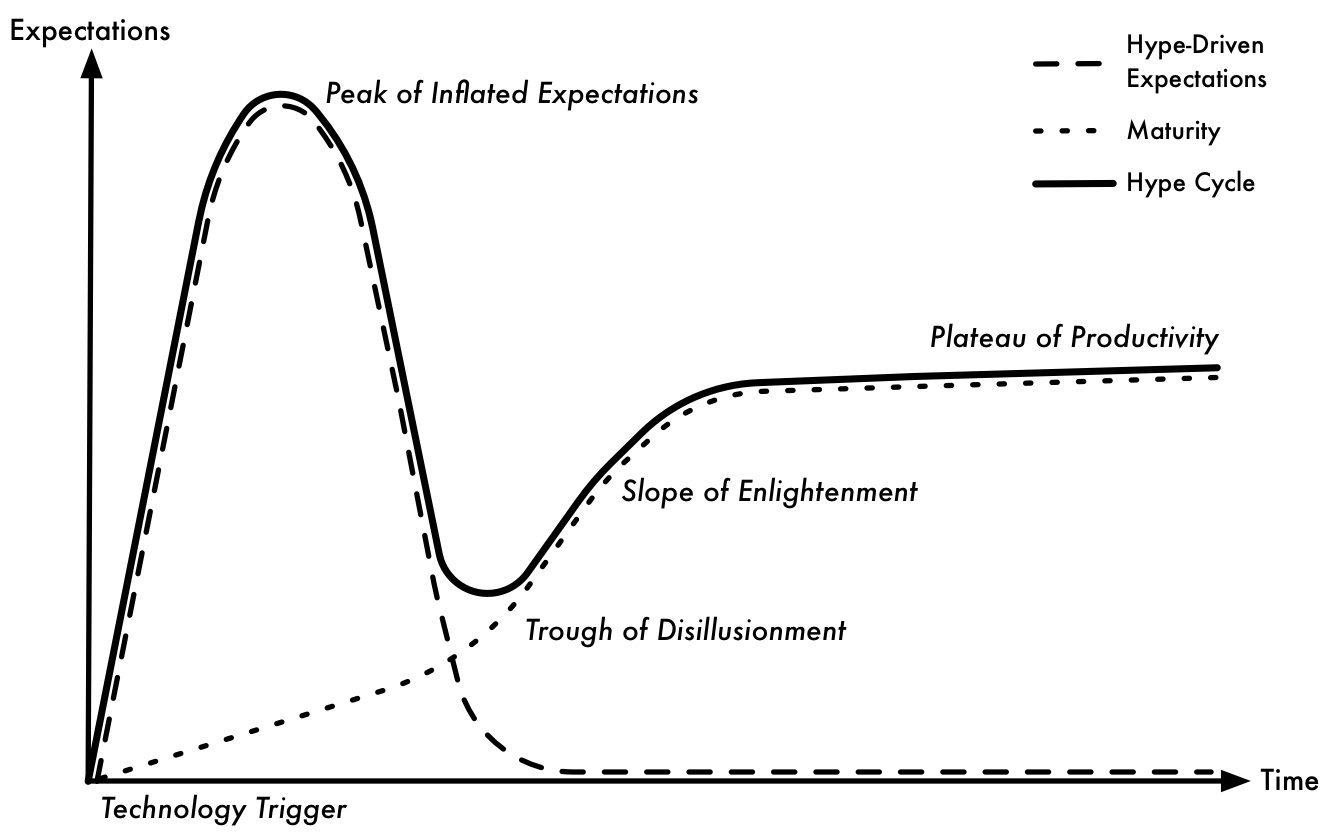}
\caption{The Gartner hype cycle \cite{Fenn2008}.}
\label{hype_cycle}
\end{center}
\end{figure}
Each year, Gartner publishes around $100$ hype cycles with insight on about $1,900$ different technologies. The framework has gained substantial attention from practitioners to facilitate strategic investment decisions, and recently from academic scholars. 

\vspace{0.4cm}
\textit{How Does the Hype Cycle Work?}

According to Gartner's hype cycle framework \cite{Fenn2008}, emerging technologies traverse a sequence of five distinct phases (see Figure \ref{hype_cycle}) that begins with over-enthusiasm for a new technology. 

\begin{itemize}

\item[$\bullet$] In the first phase, referred as the \textit{Technology Trigger}, breakthrough research and technology innovation is made. It  attracts significant interest, even before proof of commercial viability.  

\item[$\bullet$] In the second phase, known as the \textit{Peak of Inflated Expectations}, the feedback produced by early adopters and the first successful basic implementations produce a wave of unrealistic expectations about the new technology as scientists, entrepreneurs, and investors let their imaginations run loose. At this stage, the maturity level and real capabilities of the technology innovation lie far below the raised expectations. 

\item[$\bullet$] In the third phase, the \textit{Trough of Disillusionment}, interest in the proposed innovation decreases as technical challenges manifest themselves, and the technology fails to live up to the expectations raised by the early adopters. Some of these early adopters turn the corner into a period of disillusionment, due to their negative experiences. As companies bail out, investments fail to generate returns. 

\item[$\bullet$] In the fourth phase, referred to as the \textit{Slope of Enlightenment}, the true potential and benefit of the new technology start to be better understood by a select group of true believers. These hardy individuals and organizations begin experimenting with second- or third-generation products, as they fund more pilot projects, and ultimately the global knowledge grows about how to efficiently use the new technology.

\item[$\bullet$] Finally, in the fifth phase, the \textit{Plateau of Productivity}, the benefits of the technology innovation are demonstrated and accepted, and mainstream adoption takes off. Correspondingly, the associated investments start paying off. The height of the plateau depends on the quality of the technology on the one hand, and on the market size on the other hand. 
\end{itemize}

Although the Gartner hype cycle framework has gained substantial attention from practitioners for its explanatory power and predictive value, and likewise from academic scholars, the framework's theoretical underpinnings, to which this paper addresses, have been barely studied in the existing literature. 

We propose a model to quantify the behavior of interacting agents on a collective scale, as a tool to help understand the early diffusion of technological innovation. Following \cite{Fenn2008}, the hype cycle model aims at modeling the successive phases a new technology goes through, and it can be decomposed into two distinct and merged components, namely the \textit{Hype-Driven Expectations} and the \textit{Technology Maturity} (see Figure \ref{hype_cycle}). While the first component is the result of social interactions, the second one represents the pure technical performance and maturity level of the newly introduced technology. In this paper, we focus on the human component (\textit{i.e.}, the social interactions)  of the hype cycle. We therefore pay attention to the microscopic interactions that lead to the initial over-enthusiasm for the new technology, which later is followed by disillusionment after the first implementations do not meet these overly promising expectations. 


\vspace{0.4cm}
\textit{Multi-Agent Modeling}

In this paper, we interpret the hype cycle phenomenon as a  macroscopic  temporal  pattern  resulting from the microscopic interactions of a large collection of autonomous decision-making agents. Hype oscillations can be viewed as an opinion dynamics phenomenon. Hence, adopting a multi-agent approach becomes natural. Such a general framework has been pioneered in $1936$, when John Maynard Keynes described the equity markets as driven by speculators' expectations of what average opinion expects the average opinion to be\footnote{\cite{Keynes1936}: \textit{"It is not a case of choosing those which, to the best of one's judgment, are really the prettiest, nor even those which average opinion genuinely thinks the prettiest. We have reached the third degree where we devote our intelligence to anticipating what average opinion expects the average opinion to be. And there are some, I believe, who practice the fourth, fifth and higher degrees."}}.

\noindent In the present approach, we view the collective opinion underlying the progressive hype generation as the result of mutual information exchanges between technology experts. These experts mutually share either their enthusiasm ${\cal E}$ or conversely their skepticism ${\cal S}$ regarding a new technology emerging on the market. Their ${\cal E}$ versus ${\cal S}$ opinions are updated according to  gradually refining set of inquiries.  Initially, ${\cal E}$  dominates since potential drawbacks and/or flaws have not yet been unveiled. Later, thanks to the refinement of the information-gathering process, skepticism grows and ultimately dominates, producing the \textit{Trough of Disillusionment}. The  steady refinement of the information-gathering together with the mutual information sharing processes will be stylized via a stochastic nonlinear dynamics which allows for an  explicit analytic treatment. In our modeling approach, hype oscillations result from the presence of a delay into the dynamics, which is itself generated by the time required to update the economic agents' opinions.

\vspace{0.2cm}
\textit{State of the Art}

In both the academic and the business worlds, the concept of hype is commonly  used to determine the evolution and maturity level of a newly introduced technology \cite{Dedehayir2016}. A general framework was presented in \cite{Sornette2008} to model social "bubbles" and their aftermath as the consequence of collective over-enthusiasm. According to the author, this phenomenon is based on positive feedback loops of reinforcing behaviors that are linked to imitation, herding, and self-organized cooperation, which ultimately lead to the development of endogenous instabilities. The \textit{Social Bubble Hypothesis} is proposed as \textit{levers of innovation}, describing how social interactions between enthusiastic agents result in the diffusion of technological advances.

As stated in \cite{Borup2006}, the emergence of the hype cycle phenomenon is mainly due to the combination of two factors. On the one hand, the actual maturity of the technology directly affects the shape of the hype formation. On the other hand, early high-rising expectations about the technology are due to complex social interactions, and they are often spread through spoken and written words. These initial expectations can be defined as \textit{"real time representations of future technological situations and capabilities"} \cite{Borup2006}. Consultancy firms, in particular Gartner Inc. \cite{Fenn2008}, have long been using this framework to provide advice in strategic decision-making. In \cite{Dedehayir2016}, it is highlighted that the Gartner hype cycle has long been established as an empirical framework to describe the different phases of development and maturity of a new technology. Accordingly, it is widely used by practitioners in determining the current stage of a technology innovation and helping them in their investment decisions. While the hype cycle tool has been proven to be effective in practice, it has not been fully discussed on a scientific ground yet. However, the hype cycle model has attracted substantial attention from scholars and has been considered in numerous research fields dealing with the introduction and adoption of new technology. These studies range from the medical domain \cite{Heading2017}, the energy sector \cite{Khodayari2017}, information systems \cite{OLeary2008}, consumer electronics \cite{Jarvenpaa2008}, and even education \cite{McPherson2015}. In the present contribution, we adopt a generic point of view, the goal of which is to understand the nature of the multi-agent mechanisms responsible for the formation of the hype cycle. While based only on a restricted number of control parameters, our modeling nevertheless allows a fitting to generic hype patterns.


An attempt to describe the hype cycle quantitatively has been recently proposed in \cite{Silvestrini2017}. To that aim, a mathematical approach based on a probabilistic master equation is introduced, where all the required rate parameters are estimated from data by a fitting procedure. The basic difference between the present contribution and \cite{Silvestrini2017} is that here the hype phenomenon is viewed as emerging from microscopic human interactions. In \cite{Silvestrini2017}, a unified framework is proposed to fit to any existing macroscopic hype pattern, but the very origin of the hype phenomenon is not addressed. Similar bubble-like dynamic behaviors are also thoroughly discussed in closely related areas.
In particular, in the context of the stock market, a multi-agent model is proposed in \cite{Harras2011} to describe the spontaneous development of bubbles and subsequent crashes. The interacting agents base their behavior on various type of information sources: public information, peers, and private information. It is shown that bubbles originate from an initial random streak of positive news, which is then relayed and magnified by a nonlinear feedback mechanism.

\vspace{0.4cm}
\textit{Structure of the Paper}

The rest of the paper is organized as follows.
In Section \ref{Section_Collective}, we introduce the multi-agent dynamics that model the collective emergence of the hype cycle. In Section \ref{QS_Representation}, we draw an analogy between the proposed microscopic hype cycle behavior and a queuing system with feedback mechanism. The queuing system is roamed by agents who, based on their individual experience (\textit{i.e.}, the past waiting time before receiving service), decide on their upcoming routing  decision (\textit{i.e.}, revisit or leave the network). Based on this queuing analogy, we analytically compute the characteristics of the considered hype cycle dynamics in Section \ref{Analytical}. In Section \ref{Detection},  we show how to specifically implement a general detection algorithm to forecast the onset of the collective disinterest. Finally, concluding remarks and perspectives are given in Section \ref{Conclusion}. 

 

\section{Collective Opinion Dynamics and Emergence of the Hype Cycle}\label{Section_Collective}

As highlighted in \citep{Silvestrini2017}, innovation hype cycles are macroscopic patterns depicting the time-dependent collective opinion of a society of autonomous, mutually interacting, decision-making agents. Along these lines, we propose a multi-agent model that synthesizes the collective opinion oscillation (hype like) dynamics of a population of potential adopters (PAs) of an innovation. The opinion of the PAs is formed by gathering publicly available information.

Specifically, we assume the existence of an abstract information pool ${\cal P}$, the content of which is openly available. The facility ${\mathcal P}$ stores  abstract {\it information tokens} (ITs), denoted by $Q(q, \tau; t)$, and the average incoming rate of ITs is written as  $\lambda$.  Three internal variables characterize each IT, namely \textit{(i)} the running time $t$, \textit{(ii)} an opinion binary variable $q$ (${\cal E}$ standing for \textit{enthusiast}, ${\cal S}$ for \textit{skeptic}), and \textit{(iii)} the storage waiting time $\tau \in \mathbb{R}^{+}$ spent by the IT when stored  in  ${\mathcal P}$. Upon entrance into ${\mathcal P}$, the ITs always exhibit the internal states  $q={\mathcal E}$ and $\tau=0$. The updating rule for the binary variable $q$ will be specified below. At any time $t$, the collective opinion of the swarm of PAs is quantified by the number ${\cal N}(t)$ of ITs stored in ${\mathcal P}$ (\textit{i.e.}, the larger is  ${\mathcal N}(t)$, the stronger is the incentive for adopting the new technology). Following this representation, a hype oscillation therefore corresponds to a peak followed by a trough of ${\mathcal N}(t)$.




\noindent In parallel to the IT storage pool ${\cal P}$, we consider a collection of $\mathcal{B}$  information gathering experts (GEXs). These GEXs stylize the role of opinion leaders who ultimately influence the collective decisions of the PAs by sharing their judgment about the new innovation. The task of the GEXs is to extract, one by one\footnote{Accordingly, at most $\mathcal{B}$ ITs can be simultaneously extracted from ${\mathcal P}$, and one assumes that as long as ITs are stored in ${\mathcal P}$, no GEX remains idle.}, ITs from the information pool ${\mathcal P}$. The extraction policy consists in systematically selecting the IT exhibiting the highest waiting time $\tau$. After extraction, the GEX reads the $q$-state posted by the IT. We assume that the IT extraction and reading process requires a random time with average $\mu^{-1}$ and that $\mathcal{B}\mu > \lambda$, thus ensuring that, on average, the outgoing flow of ITs exceeds the externally incoming one, thus precluding from instability. 

\vspace{0.5cm}
\noindent {\it IT Dispatching Rule}

\noindent Depending on the $q$-state exhibited by an extracted IT, two routing alternatives are adopted:


\vspace{0.2cm}
\noindent \textit{(1)} When $q = {\mathcal E}$ (\textit{enthusiastic}), a new IT with internal variables $Q({\mathcal E}, 0; t)$ is fed back to ${\mathcal P}$ (\textit{i.e.}, the internal variables are updated with $\tau =0$ and $q= {\mathcal E}$).

\vspace{0.2cm}
\noindent \textit{(2)} When $q = {\cal S}$ (\textit{skeptical}), the IT is permanently discarded from the system and hence the content of ${\mathcal P}$ is depleted by one unit.

\vspace{0.5cm}
\noindent {\it Updating  of the Internal Variable $q$}

\vspace{0.2cm}
\noindent  The transitions  ${\mathcal E} \mapsto {\mathcal S}$ are random and occur at a rate monotonously increasing with the sojourn time $\tau$. We assume that there is no transition ${\mathcal S} \mapsto {\mathcal E}$.

   
 
\vspace{0.5cm}
 \noindent {\it Highlighting the Salient Modeling Features} 

\vspace{-0.12cm}
\begin{itemize}
\item[$\bullet$] By requiring the ${\mathcal E} \mapsto {\mathcal S}$ transition rate to be monotonously increasing with the sojourn time  $\tau$ of the ITs in ${\mathcal P}$, we implicitly assume that during their storage in ${\mathcal P}$, the ITs follow an information maturation process. This models the fact that, while stored in ${\mathcal P}$, ITs steadily scavenge information from their surrounding and hence are potentially better able to unveil potential shortcomings of the innovation under evaluation. This is stylized by a $\tau$-dependent enhancement rate of the transitions ${\mathcal E} \mapsto {\mathcal S}$.
 
\item[$\bullet$] By requiring the GEXs to systematically extract those ITs posting the highest $\tau$, the IT feedback flow in ${\mathcal P}$ is reduced to the minimum, and hence the $\mathcal{N}(t)$-level is minimized. Accordingly, when monitoring $\mathcal{N}(t)$, the PAs are guaranteed to optimize the available information maturity.
 
\item[$\bullet$] Restricting the number of GEXs to $\mathcal{B}$  effectively stylizes the information saturation mechanism.


\item[$\bullet$] The agent behavior under study shares some similarities with the \textit{"follow the crowd or avoid it"} principle exposed in \cite{Hassin2003} in the context of queuing theory. However, the presence of a delay in the decision-making precludes here from the emergence of a (time-independent) stationary state.
\end{itemize}

\section{Alternative Representation of the Dynamics - Feedback Queuing System with Impatient Customers and  FIFO Service Discipline}\label{QS_Representation}

\begin{figure}[htbp]
\begin{center}
\includegraphics[height=6cm]{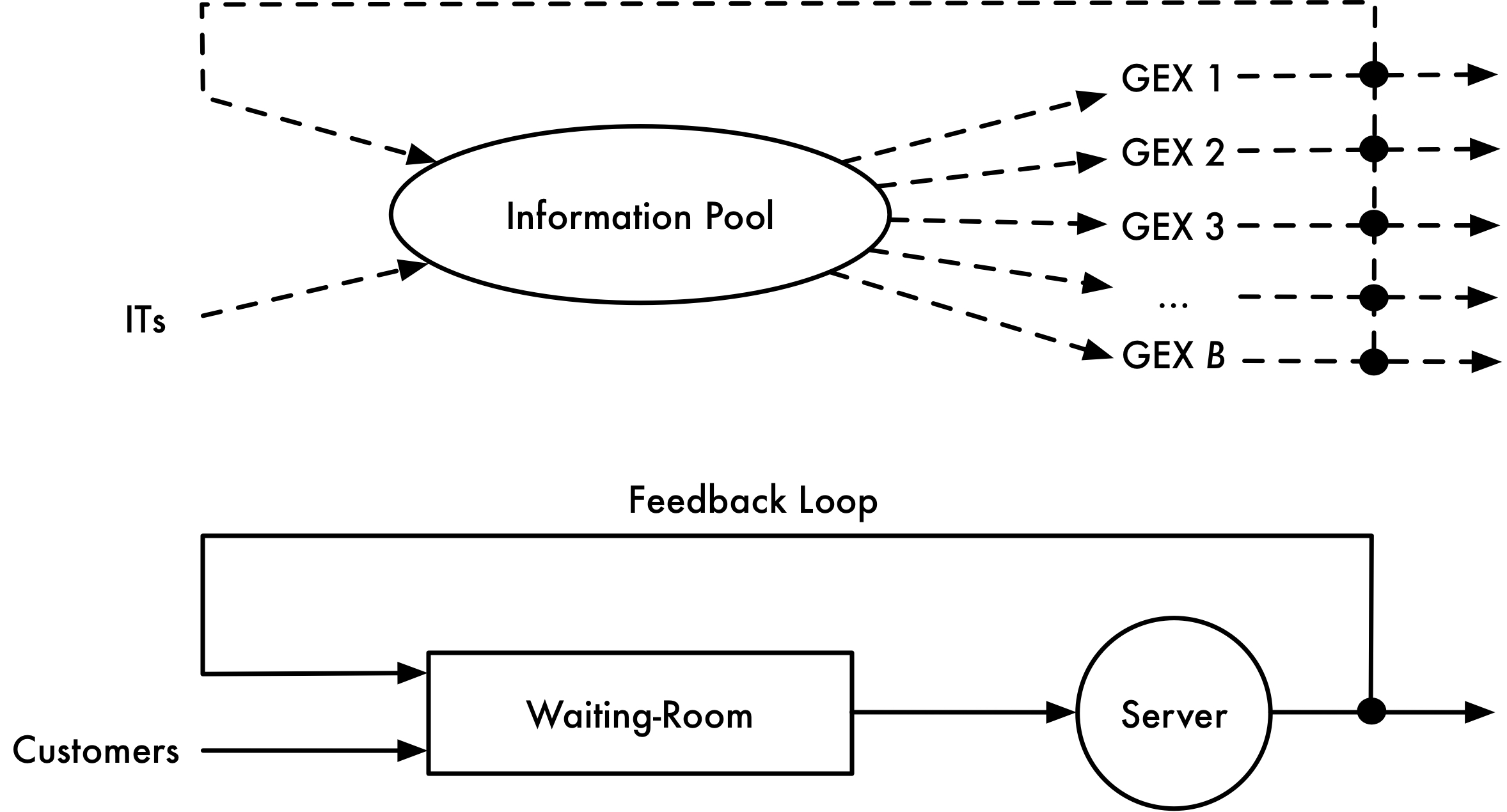}
\caption{Collective information flow leading to the emergence of the triggering phase of the hype cycle (above), and corresponding abstract feedback QS representation (below).}
\label{SB}
\end{center}
\end{figure}

\noindent As sketched in Figure \ref{SB}, 
the agent mechanisms introduced in Section \ref{Section_Collective} can be naturally summarized by considering a $\mathcal{B}$-multiple server queuing system (QS) with a feedback loop. An external random flow of incoming customers (corresponding to the ITs) feeds a QS composed of $\mathcal{B}$ parallel servers. The incoming ITs are stored inside a waiting room (here denoted by ${\cal P}$), and they are subsequently served in parallel by a set of multiple servers (representing the GEXs). In this abstract QS, the ITs in circulation behave effectively as impatient customers who, after being served, autonomously decide either to use the feedback loop or to definitely leave the system. Their individual routing decisions are taken on the basis of their last experienced waiting time in the QS (denoted here by the internal variable $\tau$). If $\tau$ exceeds a critical threshold $T$ (\textit{i.e.}, the customer's patience), the customer chooses to leave the system. The service discipline is {\it First-In-First-Out} (FIFO), which corresponds to the GEXs always serving the IT posting the highest $\tau$.

The QS analogy continues by assuming that the random times between two successive exogenous incoming ITs is drawn from a stationary probability density $f(t)dt$ with average $1/\lambda:= \int_{0}^{\infty} tf(t) dt$. Similarly, the random time interval required to read the internal state of an IT is drawn from a stationary probability law $g(t)$ with average $1/\mu:=  \int_{0}^{\infty} tg(t) dt$. For simplicity, we assume in the sequel that $\lambda^{-1}  >  (\mathcal{B}\mu)^{-1} $, and hence the average service time is shorter than the average inter-arrival time, thus ensuring queue stability. Finally, $h(\tau)$ denotes the hazard rate function characterizing the ${\cal E}\mapsto {\cal S}$ transitions\footnote{$h(\tau)$ is assumed to be identical for all ITs.}. 

\section{Hype Cycle - Analytical Derivation of the Dynamics}\label{Analytical}

\noindent To analytically calculate the dynamical queue content ${\mathcal N}(t)$,  we now make the following simplifying assumptions\footnote{As it should later become intuitively clear, an analytical discussion remains feasible even when some of these simplifying assumptions are relaxed.}:
\begin{itemize}
\item[\textit{(i)}]  \quad $h(\tau) = \delta (\tau -T)$
\item[\textit{(ii)}]  \quad $T > >\max\left\{1/ \mu, 1/ \lambda \right\}$
\end{itemize}
\noindent Assumption \textit{(i)} reflects a deterministic routing threshold and Assumption \textit{(ii)} allows us, thanks to the strong law of large numbers, to  focus  on  a purely deterministic reasoning. At time $t=0$, we assume ${\mathcal N}(0)=0$ (\textit{i.e.}, the information pool ${\mathcal P}$ is initially empty, as no available judgment nor expectation are available before the \textit{Technology Trigger}). During short transients,  all incoming ITs are fed back to ${\cal P}$ since we systematically have $\tau<T$ and so all ITs post  $q={\mathcal E}$ (this is a direct consequence of assumption \textit{(i)}). Accordingly, the content  ${\mathcal N}(t)$  of ${\cal P}$ monotonously  increases  at the average rate $\lambda$ (\textit{i.e.}, the exogenous rate of the incoming ITs, since none leaves the system). This stylizes the building of the hype pattern. As time proceeds, an increase of $\tau$ (\textit{i.e.}, the waiting time of the stored ITs) arises together with the increase of ${\mathcal N}(t)$. Once ${\mathcal N}(t)$ reaches a critical threshold, part of the stored ITs were stored long enough so that $\tau >T$. This in turn implies that these  ITs exhibit $q$-transitions from ${\mathcal E}$ to ${\mathcal S}$ and will subsequently be ejected from the system. Remember that GEXs always  extract from ${\cal P}$ the ITs posting the largest  ${\mathcal \tau}$, which effectively implements a FIFO discipline. So waiting ITs, which undergo ${\mathcal E} \mapsto {\cal S}$ transitions, form a row-cluster waiting before being extracted from ${\mathcal P}$. With a given time delay, this row-cluster reaches service and is followed by a spontaneous collective ejection leading to a swift depletion of the content ${\mathcal N}(t)$ of ${\cal P}$. This stylizes the \textit{Trough of Disillusionment} of the hype cycle pattern. Note that before being extracted, waiting ITs exhibiting $q={\mathcal S}$ remain temporarily stored in ${\mathcal P}$ until their definitive ejection. The elapsed time between the ${\mathcal E} \mapsto {\mathcal S}$ transitions and the actual extraction time implies  a delayed response dynamics, which is responsible for  the sudden decrease of ${\mathcal N}(t)$. 

\begin{figure}[htbp]
\begin{center}
\includegraphics[height=5.5cm]{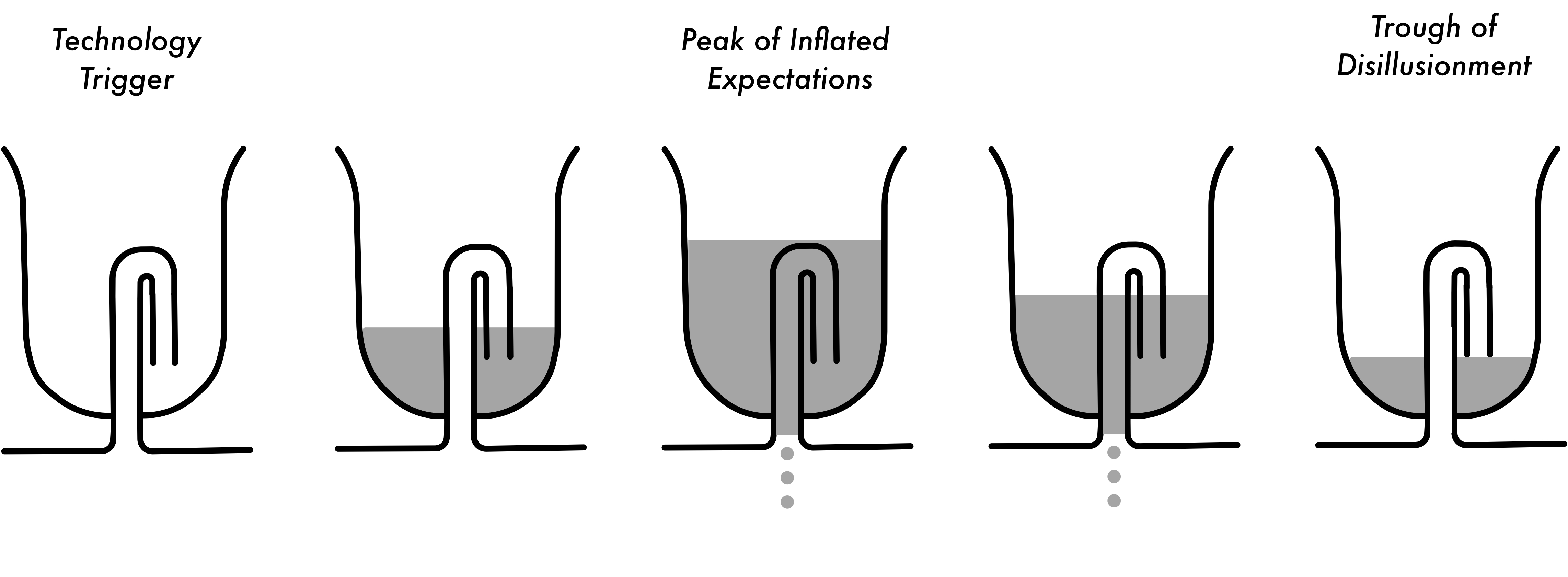}
\caption{{\bf The Tantalus or Pythagorean glass device}. In the three leftmost panels, liquid is accumulating (at rate $\lambda$) into the glass, which corresponds to expectations inflating until they reach the \textit{Peak of Inflated Expectations}. In the two rightmost panels, liquid is escaping from the glass (\textit{i.e.}, siphoning of the glass content, at rate $\lambda-\mu$), representing the phase where disillusionment prevails until the {\it Trough of Disillusionment} is reached. The siphoning rate directly depends on the diameter of the internal tube. The distance between the bottom of the glass and the top of the internal tube is directly  proportional to $T$. Observe that for large values of $T$, any fluctuations in the filling rate will be averaged during the filling cycle (\textit{n.b.}, this is an analogical manifestation of the strong law of large numbers), thus enabling a deterministic approach.}
\label{HYPEGART}
\end{center}
\end{figure}

\noindent Intuition about the ${\cal N}(t)$-dynamics can be developed  by invoking the hydrodynamic analogy  with  the  flow arising inside a Tantalus glass device (\textit{n.b.}, also referred to as the Pythagorean glass), as sketched in Figure \ref{HYPEGART}.

\vspace{0.1cm}
\textit{Calibration of the Hype Cycle}

\vspace{0.2cm}
\noindent 
Only three independent  parameters are required for the model calibration, namely:
\begin{itemize}
\item[\textit{(a)}]  $1/\lambda$:  the mean inter-arrival time of exogenous ITs,
\item[\textit{(b)}] $1/\mathcal{B}\mu$: the mean service time needed to extract one IT from ${\cal P}$ when $\mathcal{B}$ servers are engaged,
\item[\textit{(c)}] $T$: the critical waiting time, which triggers the ${\cal E} \mapsto {\cal S}$ switches.
\end{itemize}
These three control parameters enable replication of the three variables listed in \cite{Vanlente2013} for characterizing the shape and size of a hype pattern, namely \textit{(i)} the degree of enthusiasm during the positive hype leading to the \textit{Peak of Inflated Expectations}, \textit{(ii)} the degree to which enthusiasm breaks down in the negative hype leading to the \textit{Trough of Disillusionment}, and \textit{(iii)} the overall length of the hype pattern.

\noindent To simplify the discussion, we focus on  $\mathcal{B}=1$ from now on, meaning that the full service capacity is used as long as the storage is not empty\footnote{Note that provided $T$ is large enough, full service capacity will always be realized and, in this case, we have to simply substitute $\mu \mapsto \mathcal{B} \mu$ in the $ \mathcal{B}=1$ results given in Eq.(\ref{HYPCYCLOR}.)}. For large parameter $T$, the dynamics of ${\mathcal N}(t)$ becomes quasi-deterministic and the resulting hype oscillation exhibits a saw-tooth shape fixed by the three above calibrating parameters. Explicit expressions for the time duration $H$ and amplitude $\Delta$ of the hype cycle read as \cite{Gallay2009}\footnote{In \cite{Gallay2009}, focus is put on describing a perpetual periodic queue content. In the present application, a single cycle of such periodic dynamics is considered.}:

\begin{equation}
\label{HYPCYCLOR}
\left\{ 
\begin{array}{l}
 H = T \left[ 2 + {\lambda \over \mu - \lambda } + {\mu - \lambda \over \mu}\right]
  \\ \\
 \Delta = T \mu
\end{array}
\right.
\end{equation}

\noindent The above stylized hype-dynamics suggest hence a possibility to propose, with only three calibrating parameters, a rough classification encompassing the hype patterns arising in a multitude of different innovation domains.
\section{Detection of the Peak of Inflated Expectations}\label{Detection}

\noindent Having proposed a dynamic model for the hype cycle generation, the next natural and  truly relevant issue is to   quickly detect the onset of the negative hype leading to the \textit{Trough of Disillusionment} (\textit{e.g.}, to prevent from bad investments). The objective consists in implementing the {\it quickest feasible detection}\footnote{More precisely stated: detect as soon as possible, without triggering false alarms.} of a drop of the IT arrivals in ${\mathcal P}$. This type of detection issues  occupies a prominent place in the statistics literature for several decades. For the situation at hand, the problem can be reduced to a so-called  {\it Poisson disorder} or {\it Wiener disorder} problem, where either the rate of a Poisson process or the drift of a Brownian motion is suddenly altered at an unknown random time $\theta \in \mathbb{R}^{+}$. In the sequel, we will make use of the general methodology developed  by  A. Shiryaev in \cite{Shiryaev2010}, and hence we focus explicitly on the Wiener disorder problem.

\noindent For the hype-dynamics developed above, the considered problem aims at detecting as soon as possible, and ideally without false alarms, the onset of the \textit{Trough of Disillusionment}. To this goal, and in order to adapt the detection scheme exposed in \cite{Shiryaev2010}, we decompose our analysis into two steps: 
\begin{itemize}
\item[\textit{(1)}] Construct a diffusive approximation for the IT stochastic filling process of ${\mathcal P}$.
\item[\textit{(2)}] Apply the Shiryaev's method \cite{Shiryaev2010}  to detect  the change of the drift of the diffusion  process (\textit{i.e.}, detect the onset  of the \textit{Trough of Disillusionment}).
\end{itemize}

\vspace{0.2cm}
\noindent \textit{(1)  Diffusive Approximation for the Filling Process of ${\mathcal P}$}

\vspace{0.2cm}
We need to construct a stochastic differential equation (SDE) of the form:
\begin{equation}
\label{APPROSDE}
dY(t) = \mu  dt + \sigma dW_t,
\end{equation}
\noindent where $dW_t$ is a standard White Gaussian Noise process, and $\mu$ and $\sigma$ are respectively the drift and the variance, which depend on the arrival and service flows of the underlying QS. The diffusion process as written in  Eq.(\ref{APPROSDE}) must approximate the IT filling process of  ${\mathcal P}$. It is therefore the net inflow balance resulting from the superposition of three (generally not independent) processes, namely: the incoming and the feedback flows which fill ${\cal P}$, from which the flow due to service. In the early filling phase however, all ITs are fed back into ${\cal P}$, hence the average filling rate is simply given by:
$$\underbrace{(\lambda - \mu)}_{{\rm \,\, incoming \,\, -\,\,  service}}+ \quad \underbrace{\mu}_{{\rm feedback}}= \quad \lambda.$$
\noindent While the drift in Eq.(\ref{APPROSDE}) is obvious  to calculate, a refined characterization of $\sigma$ is  more  cumbersome, due to  the superposition of several (generally dependent) renewal processes. Indeed, $\sigma$ depends on both the variance of the incoming flow and of the service time. Note however that the feedback loop alone does not add any extra randomness, and hence fluctuations of the IT inflow  are only due to the inter-arrival and service times, implying that we can simply use the diffusion approximation developed  for $G/G/1$ queuing systems\footnote{The notation $G/G/1$ indicates that both the inter-arrival and the service times obey general probability laws.}, as exposed in \cite{Newell1971} or \cite{Czachorski2009} (Section II):
$$
\sigma^{2} =  \left[ \lambda C_{A}^{2} + \mu C_{S}^{2}\right] = \left[ \sigma^{2}_{A} \lambda^{3}+ \sigma^{2}_S \mu^{3} \right] , 
$$
where $C_{A}^{2}$ ($\sigma_{A}^{2}$) and $C_{S}^{2}$ ($\sigma^{2}_{S}$) respectively stand for the coefficients of variation (variances) of the inter-arrival and service times. 

\vspace{0.5cm}
\noindent \textit{(2) Detection of the Onset Towards the Trough of Disillusionment}

\begin{figure}[htbp]
\begin{center}
\includegraphics[height=8cm]{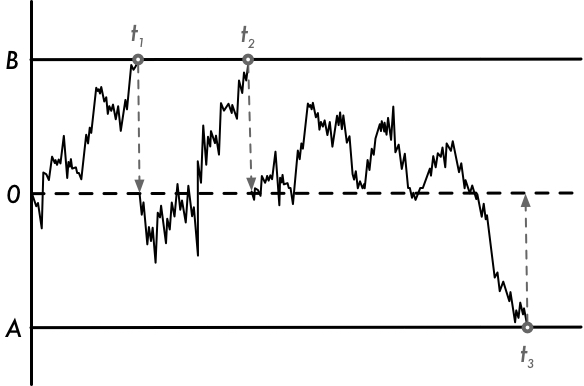}
\caption{Multistage Wald's detection method, as exposed in \cite{Shiryaev2010}. When level $B$ is reached, at time $t_1$, the process $Z_t$ returns to $0$. Just after time $t_1$, the observed process $Z_t$ becomes $Z_t - B$. At time $t_2$, when $Z_t - B$ in turn reaches level $B$, the process returns again to $0$ and the observed process becomes $Z_t = Z_t - 2B$, and so on. When the lower bound $A$ is reached (here at $t_3$), it will be interpreted as the \textit{Peak of Inflated Expectations} and the onset of the move towards the \textit{Trough of Disillusionment}.}
\label{detection}
\end{center}
\end{figure}

\vspace{0.2cm}
\noindent From Eq.(\ref{APPROSDE}), we then  consider the following detection problem:
\begin{equation}
\label{APPROSDE11}
\left\{
\begin{array}{l}
dY(t) = \mu dt + \sigma dW_t, \, \,\,\, \qquad {\rm for} \,\,\,  0 <t <\theta, \\ \\ 
dY(t) = \rho \, \mu dt + \sigma dW_t, \qquad {\rm for}\,\,\,  t \geq \theta, \\ \\ 
Y_{0} = 0,
\end{array}
\right.
\end{equation}
\noindent where $\rho\in [0,1]$ and $\theta$ is an unknown time when drift alteration occurs. In other words, Eq.(\ref{APPROSDE11}) states that, until time $t = \theta$, the  IT average filling rate of ${\cal P}$ is equal to $\mu$, while, for $t\geq \theta$, it is dropped by an external factor $\rho$, thus signaling the \textit{Peak of Inflated Expectations} (and hence the onset of the move towards the \textit{Trough of Disillusionment}). The factor $\rho \in [0,1]$ has to be chosen by the PA according to  her risk aversion. A small $\rho$ indicates that a very sensitive detection mechanism is implemented. Conversely, for higher $\rho$, we aim at detecting a large change in drift before giving an alarm (the comparative advantages versus drawbacks will be discussed below). To directly use the results derived in \cite{Shiryaev2010}, we introduce the drift shift  $U:= Y_t + \mu \rho t$, which implies that Eq.(\ref{APPROSDE11}) can be rewritten as:
\begin{equation}
\label{APPROSDE1}
\left\{
\begin{array}{l}
dU(t) = \mu(1- \rho)  dt + \sigma dW_t, \, \,\,\, \qquad {\rm for} \,\,\,  0 <t <\theta, \\ \\ 
dU(t) =   \sigma dW_t, \qquad \qquad \qquad \,\,\,  \qquad {\rm for}\,\,\,  t \geq \theta, \\ \\ 
U_{0} = 0.
\end{array}
\right.
\end{equation}
\noindent Using Eq.(\ref{APPROSDE1}), and following the lines exposed in \cite{Shiryaev2010}, we define the stochastic process $Z_t$ as:
\begin{equation}
\label{ZZ}
Z_t = \log\left\{ {dP_{0}(t)\over dP_{\infty}(t)}\right\},
\end{equation}
\noindent where $dP_{0}(t)$ and $dP_{\infty}(t)$ are respectively the probability densities for the laws of $(U_t\mid \theta=0)$ and $(U_t\mid \theta=\infty)$. According to Eqs.(\ref{APPROSDE1}) and (\ref{ZZ}), we obtain:
\begin{equation}
\label{ZZZ}
Z_t = {{\mu (1- \rho)\over  \sigma^{2}}  U_t } - { \mu^{2} \left(1- \rho^{2} \right)\over 2 \sigma^{2}}t = \left\{ 
\begin{array}{l}
+{\mu^{2}(1- \rho)^{2} \over 2 \sigma^{2} } t +  {\mu (1- \rho) \over \sigma} W_t, \qquad {\rm for} \,\,\, \theta=0, \\ \\
-{\mu^{2}(1- \rho)^{2} \over 2 \sigma^{2} } t +  {\mu (1- \rho) \over \sigma}W_t, \qquad {\rm for} \,\,\,  \theta= \infty.
\end{array}
\right.
\end{equation}
\noindent Let $A$ and $B$ be respectively a lower bound and an upper bound for $Z_t$, that will be used in the detection method (see Figure \ref{detection}). Following  \cite{Shiryaev2010}, we re-scale the time as:
\begin{equation}
\label{RESCA}
t \mapsto \tau = {\mu^{2}(1- \rho)^{2} \over 2 \sigma^{2} } t,
\end{equation}
\noindent where  $\rho$ actually denotes a  standard signal-to-noise ratio. Applying the time rescaling of Eq.(\ref{RESCA}), Eq.(\ref{ZZZ}) reduces to:
\begin{equation}
\label{SHIRRESCA}
Z_{\tau} = \left\{ 
\begin{array}{l}
+ \tau + \sqrt{2} W_{\tau}, \\ \\ 
- \tau + \sqrt{2} W_{\tau},
\end{array}
\right.
\end{equation}
\noindent which exactly matches the framework exposed in \cite{Shiryaev2010}.

\noindent Denoting $\mathbb{R}_{W}(A,B)$ and $\mathcal{T}(A,B)$ as the  {\it average detection time delay} and the {\it average time interval between two false alarms} respectively, we have, according to \cite{Shiryaev2010}:
\begin{equation}
\label{SHIRO}
\left\{
\begin{array}{l}
\mathbb{R}_{W} (A,B)= {B-A -2 \over 2} - {Ae^{(A-B)} \over 1 - e^{A}} -{A (e^{B}- e^{A})  (B -1 + e^{-B}) \over 2 \left[ A(1 -e^{B}) + B (e^{A}-1)\right]} ,
\\ \\ 
\mathcal{T}(A,B) = {A (1-e^{B}) + B (e^{A}-1))\over 1- e^{A}}.
\end{array}
\right.
\end{equation}
\noindent Consider now all possible detection bands $(A,B)$ leading to a given average time $\mathcal{T}(A,B)$ between two false alarms. Among all the pairs $(A,B)$, it is natural to select the optimal pair $(A^{*}, B^{*})$ that leads to the minimal $\mathbb{R}_{W}(A^{*},B^{*})$. Following  \cite{Shiryaev2010}, we find $A^{*}=0$, and $B^{*}$ given by:
\begin{equation}
\label{RRR}
\left\{ 
\begin{array}{l}		
\mathcal{T}^{*}=\mathcal{T} (0, B^{*})= e^{B^{*}}- B^{*} -1, \\ \\ 
\mathbb{R}_{W}(\mathcal{T}^{*}) (0, B^{*})= {1 \over \mathcal{T}^{*}} \left\{B^{*} \left[e^{B^{*}} -e^{-B^{*}}  -{B^{*} \over 2} \right] - {3 \over 2}\left(e^{B^{*}} -2 +e^{-B^{*}} \right) \right\}.
\end{array}
\right.
\end{equation}
\noindent In limiting cases, we approximately obtain:
\begin{equation}
\label{RRRB}
\mathbb{R}_{W}(\mathcal{T}^{*}) (0, B^{*})= \left\{ 
\begin{array}{l}
 \log (\mathcal{T}^{*}) - {3 \over 2} + {\cal O}\left[ {1 \over \mathcal{T}^{*}} l\log^{2}(\mathcal{T}^{*}) \right],\qquad \,\,{\rm for }\quad  \mathcal{T}^{*} \rightarrow \infty, \\ \\
 {5 \over 6 }\, \mathcal{T}^{*}+ {\cal O}((\mathcal{T}^{*})^{2}), 
\quad \quad \quad \qquad \qquad \qquad {\rm for } \quad  \mathcal{T}^{*} \rightarrow 0.
\end{array}
\right.
\end{equation}
\noindent Finally, returning  to the original scales by using  Eqs.(\ref{RESCA}) and (\ref{RRRB}), we obtain:
\begin{equation}
\label{FINOS}
\left\{ 
\begin{array}{l}
 \widehat{\mathcal{T}}^{*} =  {\sigma^{2}  \over \mu^{2}(1 - \rho)^{2}  }\mathcal{T}^{*} = { \left[\lambda C^{2}_{A} + \mu C_{S}^{2}\right]  \over \mu^{2} (1 - \rho)^{2} }\mathcal{T}^{*} , \\ \\ 
 \mathbb{R}_{W}\left(\widehat{\mathcal{T}}^{*}\right) = \left\{ 
\begin{array}{l}
 \log \left(\widehat{\mathcal{T}}^{*}\right) - {3 \over 2} + {\cal O}\left[ {1 \over \widehat{\mathcal{T}}^{*}} l\log^{2}\left(\widehat{\mathcal{T}}^{*} \right)\right],\qquad {\rm for }\quad  \widehat{\mathcal{T}}^{*} \rightarrow \infty, \\ \\
 {5 \over 6 }\, \widehat{\mathcal{T}}^{*}+ {\cal O}\left(\left(\widehat{\mathcal{T}}^{*}\right)^{2}\right), 
\quad \quad \quad \qquad \qquad \qquad {\rm for } \quad  \widehat{\mathcal{T}}^{*}  \rightarrow 0.
\end{array}
\right.
\end{array}
\right.
\end{equation}

\vspace{0.2cm}
\noindent In view of Eqs.(\ref{RRR}), (\ref{RRRB}), and (\ref{FINOS}), the following observations can be drawn. Eq.(\ref{FINOS}) quantifies the intuitive fact that imposing a very large average time between two successive false alarms $\mathcal{T}(A, B)$ leads in parallel to accepting a large delay before detection. Indeed, the requirements of simultaneously having large average times between false alarms and  short average detection times are conflicting issues. The optimal trade-off should be adjusted according to predefined utility functions which depend on the risk aversion of the PAs. Strongly risk-averse PAs are likely to insist on early detection of the \textit{Peak of Inflated Expectations}, implying that they are prepared to tolerate false alarms. Accordingly, risk aversion leads to potential loss of (more risky) rewards that can be  expected  from higher hype levels. Conversely, risk-inclined PAs accept potential losses due to false alarms, and are prepared  to  accept later warning  of the occurrence of the \textit{Trough of Disillusionment}.

\vspace{0.5cm}
\noindent {\it Additional Remarks:}
\begin{itemize}
\item[\textit{(i)}] {\it Role of the drift change  sensitivity parameter $\rho$ }

From the first line of Eq.(\ref{FINOS}), we observe the  intimate  interplay that exists between the  parameters  $\sigma^{2}$, $\rho$, and $\widehat{\mathcal{T}}^{*}$. Specifically, for  fixed  $\widehat{\mathcal{T}}^{*}$, an increase in $\rho$ leads to an effective increase of $\widehat{\mathcal{T}}^{*}$. Hence, {\it ceteris paribus}, an increase in the choice of $\rho$ (\textit{i.e.}, a decrease of the detection sensitivity) corresponds  to an effective  increase of the average time between consecutive false alarms.

\item[\textit{(ii)}] {\it Utility function and optimal control} 

As fully discussed in \cite{Shiryaev2010}, the question regarding the optimality of the present cyclic detection method might be raised. This is done by formulating a stochastic optimal control problem in which the  utility function encapsulates the trade-off between the frequency of false alarms and early detection. This refined formalism leads to a slight improvement of  Eq.(\ref{FINOS}) (see Theorem 5.1 in \cite{Shiryaev2010}). 
\end{itemize}


\section{Conclusion and Perspectives}\label{Conclusion}
We have proposed a stylized multi-agent model to capture the essence of the collective temporal patterns underlying the \textit{collective mind} phenomenon leading to the emergence of hype cycles. The present work belongs to the ongoing general line of research devoted to the \textit{Social Bubble Hypothesis}, \textit{e.g.,} \cite{Gisler2018}. Since hype cycle dynamics is a phenomenon lying at the interface between qualitative and quantitative sciences, any relevant modeling approach should offer both simplicity (\textit{i.e.}, a limited number of external control parameters to get a rapid although rough calibration) and robustness (\textit{i.e.}, a generic character implying that specific modeling details can be modified without altering the overall dynamic picture). In the modeling framework proposed here, the origin of the hype phenomenon is in essence due to the presence of a delayed response in a nonlinear and dynamic decision-making process in a random environment. Due to the simplicity of the model, an analytical approach remains feasible  for a whole  class of nonlinear and random processes. This allows for intuitively appreciating the role played by the few external parameters that are controlling the shape of the hype pattern (\textit{i.e.}, the hype amplitude, the disillusionment purging rate, and the information maturation threshold). While capturing the origin of the hype phenomenon sheds light on the opinion formation processes, the central issue to be addressed is undoubtedly how to enhance the ability for potential investors to detect, as soon as possible and without error, the \textit{Peak of Inflated Expectations} and  the onset of the  \textit{Trough of Disillusionment}. To address this early warning decision-making process, we borrow directly from the long history of optimal detection problems, which offers a natural and  mathematically well-formulated framework to detect the beginning of the disinterest phase. In practice, the inherent difficulties to detect the \textit{Peak of Inflated Expectations} lie jointly on  farsightedness, flair, experience, and also on more rigorous monitoring and detection methods.  While the proposed stylized approach cannot provide an entirely foolproof method, it offers a useful complementary trail to complete the plural toolbox of investors analyzing potential investment opportunities in emerging ideas.

\bibliographystyle{abbrv}
\bibliography{bib_hype.bib}


\end{document}